\documentclass[pre,floats,superscriptaddress,showkeys,usenames]{revtex4}
\usepackage{graphicx}
 \usepackage{mathptmx}      
 \usepackage[dvipsnames]{color}

\begin{document}
\title{Thermally induced directed currents in hard rod systems}
\
\author{Fabio Cecconi}

\address{Institute for Complex Systems CNR
Via dei Taurini 19, 00182 Rome Italy }
\author{Giulio Costantini }
\address{   CNR-ISC and Dipartimento di Fisica, Universit\`a Sapienza - 
p.le A. Moro 2, 00185, Roma, Italy}

\author{Umberto Marini Bettolo Marconi}

\address{ Scuola di Scienze e Tecnologie, 
Universit\`a di Camerino, Via Madonna delle Carceri, 62032 ,
Camerino, INFN Perugia, Italy}

\begin{abstract}
We study the non equilibrium statistical 
properties of a one dimensional hard-rod
fluid  undergoing collisions and
subject to a spatially non uniform Gaussian heat-bath and periodic potential.
 The system is able to sustain finite currents when the 
spatially inhomogeneous heat-bath and the  periodic potential profile 
display an appropriate relative phase shift, $\phi$.
 By comparison with the collisionless limit, we 
determine the conditions for the most efficient transport  
among inelastic, elastic and non interacting rods. 
We show that the situation is complex as,   
depending on shape of the temperature profile, 
the current of one system may outperform the others.
\end{abstract}

\maketitle
\section{Introduction}
Recently there has been an upsurge of interest in the understanding of non 
equilibrium systems which even in the absence of an applied bias can 
generate currents.
Typical examples are the thermal ratchets and Seebeck ratchet~\cite{Reimann}, 
where an asymmetric potential and a non Gaussian noise generate 
a directed motion.\\
Several authors \cite{VdenBroeck,Costantini} showed that a class of 
geometrically 
asymmetric elastic objects undergoing some holonomic constraint and  
coupled to heat baths at different temperatures can rectify 
thermal fluctuations and thus produce work.
The absence of a time-reversal symmetry invalidates the standard 
detailed balance \cite{Risken,Gardiner85}.
The directed motion of microscopic systems of somehow different nature 
was also studied several years ago by Landauer~\cite{Landauer75,hanggi1996brownian} 
who considered a bistable potential 
with an hot heat reservoir placed at one side of the potential peak and a 
cold reservoir on the other side and predicted a directed current of 
particles toward the colder side.
This is the so-called the blowtorch effect which has been exploited in 
Refs.~\cite{Benjamin08,Izumida} to produce directed currents in periodic 
non-isothermal systems. \\
In this paper we study how the presence of interactions, such as excluded volume and inelastic collisions among Brownian particles, affects the blowtorch mechanism.
To the best of our knowledge only the case of overdamped
independent particles has been analyzed in detail~\cite{Aswaf00,Bekele}.

\section{Model}
The model consists of $N$ impenetrable hard-rods 
of mass, $m$, size $\sigma$ and position $x_i(t)$ ($i=1,\ldots,N$) 
evolving on a segment of length $L$ according to the dynamics 
\cite{cecconi2004fluid,cecconi2004inelastic}
\begin{equation}
m \frac{d^2 x_i}{d t^2}=-m \gamma\frac{d x_i}
{d t}-\frac{d V(x_i)}{d x_i} 
+\sqrt{2 m \gamma
T(x_i)} \; \xi_i(t) +\sum_j f_{ij}.
\label{kramers2}
\end{equation}
We assume cyclic boundary conditions, so that particles crossing with positive
velocity the point $x=L$ reenter at the point $x=0$ and viceversa.

Equation (\ref{kramers2}) is based on the assumption that four kind
of forces act on the rods. These are:\\
i) the frictional force, $-m\gamma d x_i/dt$, 
proportional to the velocity;\\
ii) the gradient of a time-independent spatially periodic potential 
$V(x) = V_0 \cos(2\pi x/w)$ of period $w$, tending to confine the 
particles near its minima;\\ 
iii) the stochastic driving force, mimicking the action of a 
heath bath with a spatially non uniform temperature profile,  
has an intensity $T(x) = T_c + T_h s(x)$ with 
$s(x) = (1+\tanh[\mu\sin(2\pi/w x-\phi)])/2$ a periodic 
smooth step-like function between $0$ and $1$, 
which alternates cold $T_c$ and warm $T_c + T_h$ regions of size $w$
in $[0,L]$. The values $T_c,T_h$ characterize the temperature jump 
amplitude and $\phi = 2\pi x_0/w$ determines the mutual shift 
of $T(x)$ and $V(x)$.
All temperatures are measured in units such that the Boltzmann 
constant is $k_B = 1$. As usual,  $\xi_i(t)$ is a zero mean and 
Gaussian noise with auto-correlation
$
\langle \xi_i(t)\xi_j(s) \rangle = \delta_{ij} \delta(t-s);
$\\ 
iv) finally, the term $\sum_j f_{ij}$ indicates the resultant 
of contact impulsive forces acting on $i$ due to the particles $j\ne i$.
namely, the rods experience mutual inelastic collisions occurring 
at contact $|x_{i+1}-x_i| = \sigma$.
After each collision the velocities of a pair $(i,j)$ change according to the
rule
$v_i' =  v_i - (1+\alpha)(v_i - v_j)/2$ 
and $v_j' =  v_j + (1+\alpha)(v_i - v_j)/2$, 
where the prime indicates post-collisional values and
$\alpha$ is the coefficient of restitution.\\

\section{Theory}
In  the limit $\gamma\to \infty$ and 
$m\gamma\to const$, the multiple time scale analysis of 
ref.~\cite{marconi2006nonequilibrium,Tarazona} can be extend to the present case to derive
the following evolution of the
one-particle density $\rho(x,t)$ from the Kramers equation 
for the phase-space distribution function $f(x,v,t)$  \cite{Lopez}: 
\begin{equation}
\frac{\partial \rho(x,t)}{\partial t} +\frac{\partial J(x,t)}{\partial x} =0
\label{psi0timeb}
\end{equation}
where the associated current reads:
\begin{eqnarray}
J(x,t) &=&-\frac{1}{m\gamma} \Bigr\{\frac{\partial}{\partial x} 
\Bigl[T_g(x) \rho(x,\tau)\Bigl]-F(x) \rho(x,t) 
+\frac{(1+\alpha)}{2}\rho(x,t)\Bigl[T_g(x+\sigma) g_2(x,x+\sigma)
\rho(x+\sigma,t) \nonumber\\
&& -T_g(x-\sigma)g_2(x,x-\sigma)\rho(x-\sigma,t)\Bigl] \Bigr\}
\label{psi0timeb2}
\end{eqnarray}
The first term represents the single particle contribution to the current, 
whereas the second term, non linear in the density, accounts for the
excluded volume effect.  It contains the pair correlation function 
$g_2(x,x')$ evaluated at contact, the coefficient of restitution $\alpha$
and the granular (or kinetic) temperature $T_g$ of particles related to the 
heat bath temperature by
\begin{equation}
T_g(x) = T(x)\Bigr\{1- \frac{1-\alpha^2}{2\gamma}\sqrt{\frac{T(x)}{m\pi}}\Bigl[
g_2(x,x+\sigma)\rho(x+\sigma)+ \nonumber \\
g_2(x,x-\sigma)\rho(x-\sigma)\Bigl] \Bigr\}.
\label{granTemp}
\end{equation}
In the high density limit, the above equations indicate 
that inelasticity, temperature and density itself result intimately connected 
to determine the system transport properties.
On the other hand, in the opposite limit, the latter term in 
Eq.~(\ref{psi0timeb2}) can be neglected and 
analytical expressions of current $J$ and particle density $\rho(x)$ can be
explicitly worked out. 
Non interacting system is a meaningful comparison  
as it constitutes the low-density regime extrapolation of interacting 
particle behavior. 
The non interacting system admits two types of stationary solutions: 
those corresponding to vanishing and non vanishing current $J$ respectively.
The key quantity determining the presence of a systematic 
flux is the "entropy" integral 
$$
S(x) = \int_{0}^x d\xi \frac{V'(\xi)}{T(\xi)}.
$$
Using the periodicity of the system, we obtain the stationary current $J_0$ 
\begin{equation}
J_0 = \frac{1}{\gamma}\; 
\frac{1-\mbox{e}^{S(w)}}{ac-b[1-\mbox{e}^{S(w)}]}
\end{equation}
in terms of three constants $a,b,c$ related to $S(x)$ by
\begin{equation}
a = \int_{0}^w dx \frac{\mbox{e}^{-S(x)}}{T(x)},\quad 
b = \int_{0}^{w} dx \frac{\mbox{e}^{-S(x)}}{T(x)}
\int_{0}^x d\xi \mbox{e}^{S(\xi)}, \quad
c = \int_{0}^w dx~\mbox{e}^{S(x)}.
\end{equation}
It can be easily verified that, when $V(L)=V(0)$ (zero external load) and $T(x)$ is constant or $\phi=0$, the current automatically vanishes.
It is instructive to discuss how the current of the non-interacting system 
depends on the temperature scales $T_c$ and $T_h$. 
The current, for $T_c$ fixed, grows as the temperature step $T_h$ increases, 
for the jumps over the barrier to become more probable. 
On the other hand, if $T_h$ is fixed, $J$ does not depend monotonically on 
the temperature $T_c$.
Clearly, in the limit $T_c\to 0$, the particles have a small probability to escape from the potential minima, so that the current must vanish. In the 
opposite limit $T_c \gg V_0$, the confining effect of the energy barriers 
is negligible and the current must vanish too. 
For intermediate values of $T_c$, a maximum in the current is expected at a 
$T_c$-value which runs to zero as $V_0$ is reduced.

\section{Numerical results}

The case of interacting particles is not so fortunate as not amenable 
to analytic solution, therefore our study will be based mainly on 
numerical simulations.
The picture, indeed, remains qualitatively but not quantitatively similar to that of the non 
interacting case and reveals some interesting features. \\
When considering interacting systems, two main effects come into play which 
may lead to a behavior deviating from the non interacting system.
First, the mutual repulsion between particles induces dynamical 
correlations, either promoting the exit from a potential well 
via energetic collisions or forbidding a jump towards a too crowded well.
A meaningful parameter commonly used to take into account the crowding degree 
of a granular system is the packing fraction $\eta = N\sigma/L$ ($0\leq \eta \leq 1$).  
Second, the granular temperature of an inelastic system is generally lower than the temperature of the elastic counterpart. 
Thus, thermally activated transport is 
expected to be less efficient when dissipative collisions are at work. 

One may wonder on the specific influence of the model parameters 
$T_c/V_0$, $T_h/V_0$, $\phi$, the packing $\eta$ and inelasticity 
$\alpha$ on the transport properties of the inelastic system (Inel).
For sake of shortness, here we discuss only temperature effect 
choosing $\phi=\pi$ (which optimizes the current) and we explore 
some significant regimes in $\eta$ and $\alpha$. 
Moreover we compare the results of the inelastic system with the elastic 
(El, $\alpha=1$) and non-interacting (NI, $\eta\to \sigma/L$) ones, 
by also analyzing the conditions for the efficient transport.

\begin{figure}
\begin{center}
\includegraphics[angle=0,width=8.0cm,clip=true]{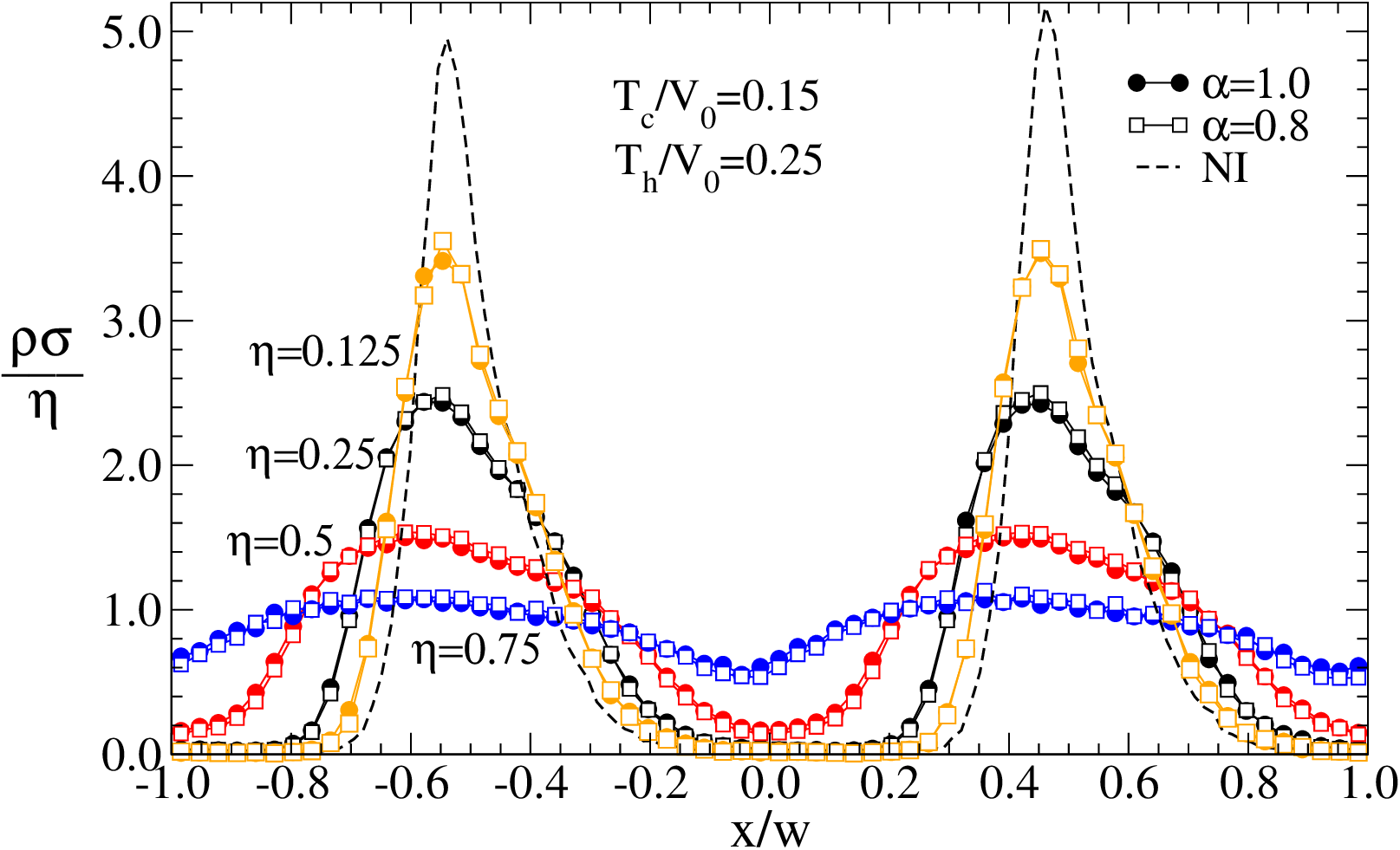}
\caption{(Color online) Density profiles of the systems at 
constant $T_c/V_0=0.15$, $T_h/V_0=0.25$ and increasing packing 
fraction $\eta=1/8$ (orange), $\eta=1/4$ (black), $\eta=1/2$ (red) and  
$\eta=3/4$ (blue). 
The dashed curve corresponds to the NI system, while the closed circles 
and open squares symbols correspond to the interacting systems with 
$\alpha=1$ and $\alpha=0.8$ respectively. 
The other parameters are the 
following: $m=1$, $\sigma=1$, $\mu = 4.0$ and $\phi = \pi$.
}
\label{fig:densy}
\end{center}
\end{figure}

\begin{figure}
\begin{center}
\includegraphics[angle=0,width=8.0cm,clip=true]{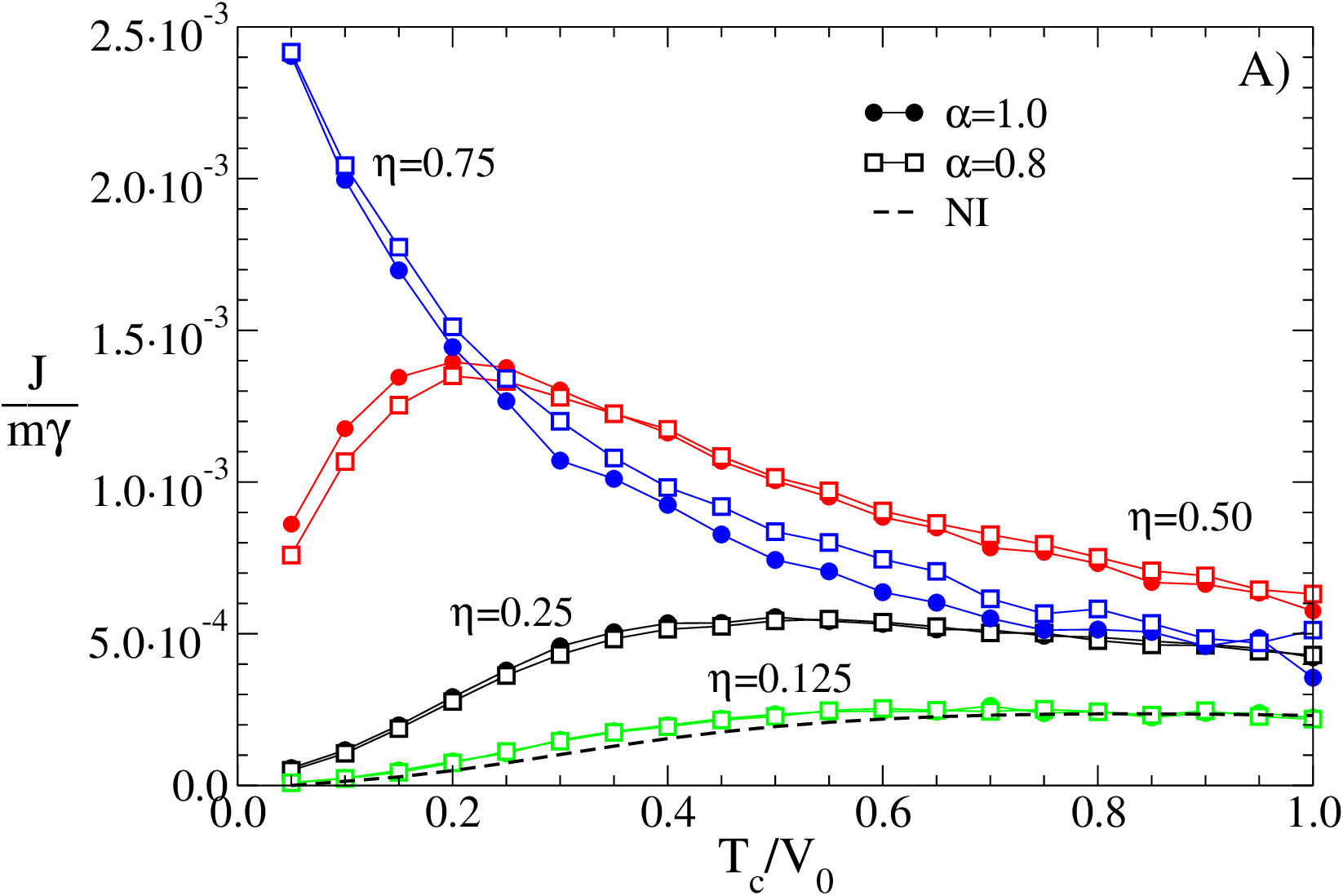}\\ 
\includegraphics[angle=0,width=8.0cm,clip=true]{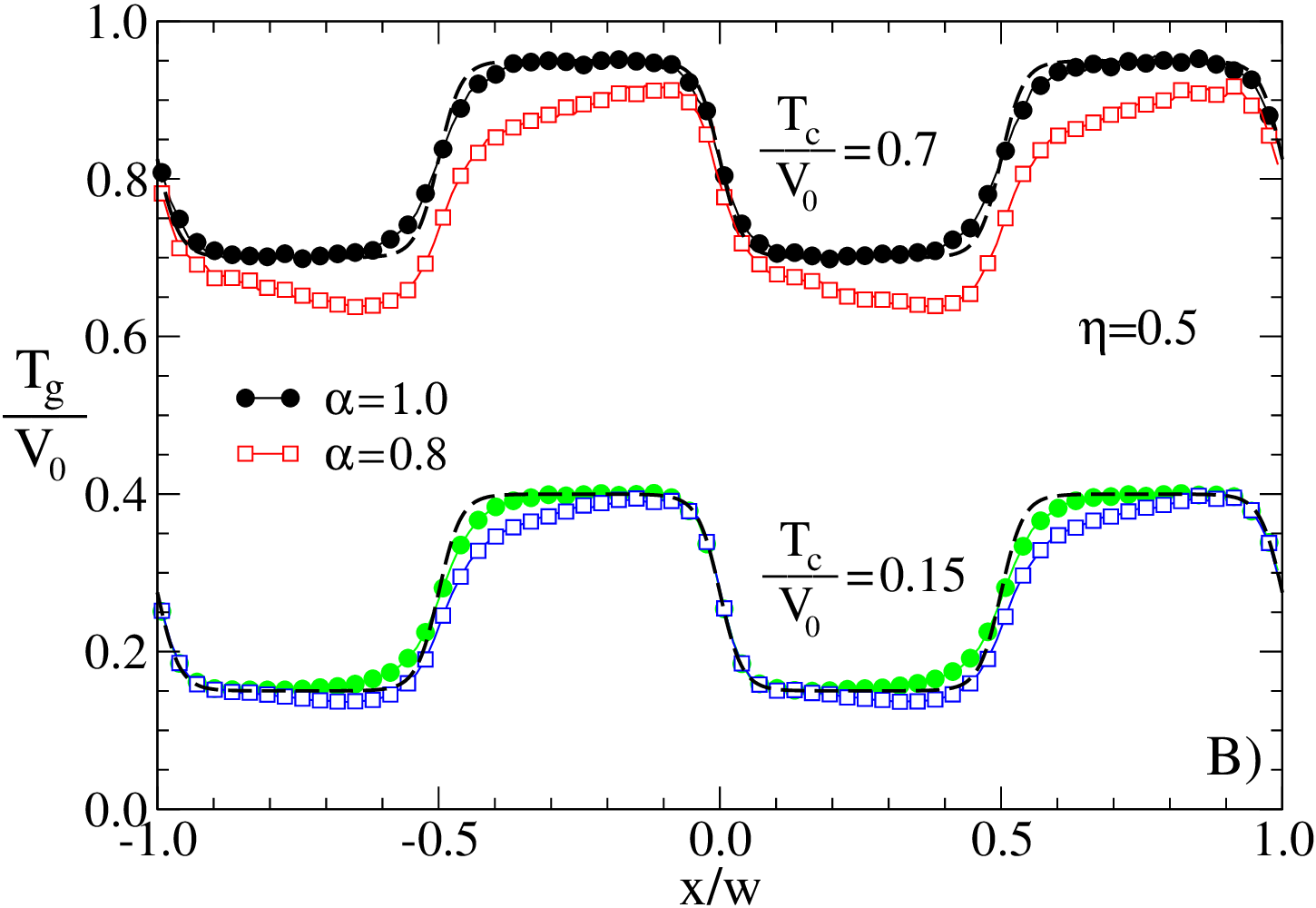} 
\caption{(Color online) A: particle current versus the rescaled temperature $T_c/V_0$ for four different values of the packing fraction 
$\eta=1/8$ (green), $\eta=1/4$ (black), $\eta=1/2$ (red) and  
$\eta=3/4$ (blue). Open and full symbols refer to inelastic and elastic 
system respectively. For sake of comparison to the low $\eta$ regime 
also the NI current is plotted (dashed curve). 
B: granular temperature profiles of the systems for two 
different value of $T_c/V_0$: $0.7$ (top) and $0.15$ (bottom). 
Open and full symbols correspond to inelastic and elastic 
system respectively. The dashed curve is the temperature profile for a NI 
system.$T_h/V_0 = 0.25$ and
the other parameters are the same as in Fig.~\ref{fig:densy}.} 
\label{fig:jcurr}  
\end{center}    
\end{figure}

It is convenient to start the discussion by considering first the stationary 
density profiles shown in Fig.~\ref{fig:densy}. The inspection of profiles, 
indeed, provides a first indication on the way particles react to parameter 
variation and how they distribute over the effective landscape generated by  
the potential and temperature profiles. 
In the NI case, with $T_h \ll V_0$ and $T_c \ll V_0$, the combined effect 
of temperature and potential profiles preferentially confines the 
particles in the 
narrow region determined by the minimum of the potential and the 
nearest colder temperature zone; in other words, 
the phase difference between the minima of $T(x)$ and $V(x)$ produces a 
sort of ``cage'' trapping the particles with small momenta. 
This feature is clearly noticeable in the $\rho_{NI}(x)$ structure of 
Fig.~\ref{fig:densy} which develops peaks in the cage-region. 
The large preferential confinement  
of the NI system is however impossible to particles with excluded volume 
interaction whose density profiles become soon broader and flatter on 
increasing the average packing fraction $\eta$, see Fig.~\ref{fig:densy}.
This corresponds to an effective decreasing of the barriers seen by the 
interacting rods and such differences in the density profiles translates 
into different transport properties. \\
We study thus the dependence of the currents $J_{NI},J_{El},J_{Inel}$ on $T_c$ 
(see Fig. \ref{fig:jcurr}) at different packings $\eta$ and $T_h/V_0=0.25$ 
fixed.
At low values of $\eta=0.125$, the current 
of the interacting systems (green symbols) behaves like the non 
interacting one (dashed black) as expected. 
Increasing $\eta$, the currents $J_{El}$ and $J_{Inel}$ while remaining 
very similar to each other, strongly deviate from the corresponding 
$J_{NI}$ obtained by multiplying for the appropriate number of 
particles, $N$, the single particle current $J_1$. 
The currents of the interacting 
systems start developing  a maximum at lower $T_c$ 
(around $T_c/V_0 \simeq 0.5$ for $\eta=0.25$ or $T_c/V_0 \simeq 0.2$ 
for $\eta=0.5$).
This behavior can be explained by means of the excluded volume effect 
that reduces the effective height of the barriers as the mean packing of 
the wells increases.
As a consequence the maximum of the current vs. $T_c$ curve is located at 
smaller temperatures with respect to the corresponding NI case.  
However, as the packing becomes sufficiently high a second effect come into 
play which changes the above scenario. 
In this regime, the effective barriers $\tilde{V}_0$ become enough small 
to be of the same order as the temperature step $T_h$ and the particles 
can escape from the potential well even if $T_c\to 0$. The currents 
$J_{El}$ and $J_{Inel}$, for $\eta=0.75$, show in fact a finite values for 
small value of $T_c/V_0$ (see blue symbols in Fig.~\ref{fig:jcurr}A). 
Moreover, increasing $T_c$, the particles become more energetic and the 
rectifying effect is reduced determining a monotonically decreasing trend 
of the currents.\\ 
Another aspect to consider is the influence of inelasticity on the transport. 
The curves in Fig.\ref{fig:jcurr}A show that, for $\eta$ sufficiently high 
(e.g. $\eta=0.5$), the inelastic system (open symbols) becomes more efficient 
than the elastic one (closed symbols) as long as $T_c/V_0>0.4$. 
On the other hand, increasing $\eta$, $J_{Inel}$ is larger than $J_{El}$ 
for all values of $T_c$ (blue symbols). This behavior can be explained 
analyzing the kinetic temperature fields of the two systems. 
In Fig.~\ref{fig:jcurr}B, we show these fields for $\eta=0.5$ and two different 
values of $T_c/V_0$.\\
While the elastic and non-interacting profiles of $T(x)$ are very close, 
the effective value of the temperature, in the inelastic case, decreases 
near the potential minima. This indicates that the ratio $T_c/\tilde{V}_0$ 
is smaller if $\alpha<1$. 
The inelastic system can thus be approximately 
considered as the elastic one, but with a lower temperature $T_c$, i.e. 
$J_{Inel} (T_c)\approx J_{El}(T_c-\delta T_c)$ for a fixed $T_c$. 
Such decreasing of $T_c$ helps to rectifying the fluctuations when the 
energy is enough for the activated thermal process, i.e. for $\eta=0.75$ 
or $\eta=0.5$ and $T_c/V_0>0.4$, or it reduces the transport when 
$T_c+T_h \ll \tilde{V}_0$ (i.e. for $\eta=0.5$ and $T_c/V_0<0.4$). \\
\section{Concluding remarks}
In this work, we have shown that contrary to intuition, one-dimensional systems with hard-core interactions can display
a more efficient ``blowtorch'' effect than a non interacting one. 
Our simulations show this efficiency inversion to be correlated to the 
resistance of hard-rods to localize in narrow regions. 
Under some conditions, the inelasticity makes the transport more efficient 
by reducing the average kinetic energy.\\

\begin{acknowledgements}

This paper is dedicated to the memory of Isaac Goldhirsch, a
great physicist who shared with us his deep understanding of the dynamics 
of granular systems.

The work of G.C. is supported by the Italian MIUR grant RBID08Z9JE.
  
\end{acknowledgements}
\bibliographystyle{unsrt}        
\bibliography{granu.bib}          
\end{document}